\documentclass[
 reprint,
 amsmath,amssymb,
 aps,
]{revtex4-1}

\usepackage{mathtools}
\usepackage{multirow}
\usepackage[export]{adjustbox}
\usepackage{graphicx}% Include figure files
\usepackage{dcolumn}% Align table columns on a decimal point
\usepackage{bm}% bold math
\usepackage{footmisc}
\usepackage[english]{babel}
\usepackage{natbib,hyperref}
\usepackage[normalem]{ulem}
\usepackage{comment}
\usepackage{xcolor}
\usepackage{subcaption} 
\usepackage{booktabs} 
\usepackage{algpseudocode}

\begin{document}

\preprint{APS/123-QED}
\graphicspath{ {./images/} }

\title{LASSE: Learning Active Sampling for Storm Tide Extremes in Non-Stationary Climate Regimes}\thanks{Corresponding Authors: Grace Jiang (gacezj@mit.edu) and Sai Ravela (ravela@mit.edu)}
\author{Grace Jiang}
\author{Jiangchao Qiu}
\author{Sai Ravela}
\address{Earth Signals and Systems Group\\
Massachusetts Institute of Technology\\
77 Massachusetts Avenue, Cambridge, MA 02139}

\begin{abstract}
Identifying tropical cyclones that generate destructive storm tides for risk assessment, such as from large downscaled storm catalogs for climate studies, is often intractable because it entails many expensive Monte Carlo hydrodynamic simulations. Here, we show that surrogate models are promising from accuracy, recall, and precision perspectives, and they ``generalize" to novel climate scenarios. We then present an informative online learning approach to rapidly search for extreme storm tide-producing cyclones using only a few hydrodynamic simulations. Starting from a minimal subset of TCs with detailed storm tide hydrodynamic simulations, a surrogate model selects informative data to retrain online and iteratively improves its predictions of damaging TCs. Results on an extensive catalog of downscaled TCs indicate a 100\% precision retrieving the rare destructive storms using less than 20\% of the simulations as training. The informative sampling approach is efficient, scalable to large storm catalogs, and generalizable to climate scenarios.
\end{abstract}

\maketitle

\section{Introduction}
\begin{figure}[htb!]
    \includegraphics[width=0.4\textwidth]{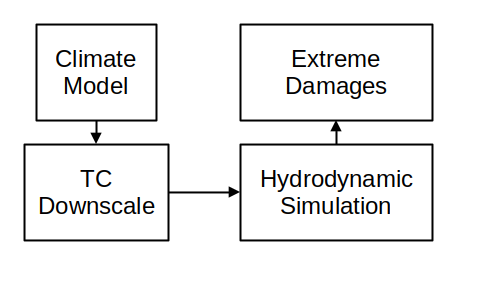}
    \caption{The standard workflow to quantify TC-induced storm tide hazards includes downscaling a climate model followed by expensive hydrodynamic simulations that make hazard assessments of extremes computationally intractable.}
    \label{fig:wkflw_before}
\end{figure}

Tropical cyclones (TCs) are among the deadliest and costliest natural disasters worldwide~\cite{needham2015review}. These storms bring strong winds and torrential rain and often cause compound flooding, wreaking havoc on infrastructure and livelihoods. Climate change exacerbates these hazards by increasing the frequency and intensity of TCs and driving sea-level rise. For example, in Bangladesh, the polder embankment systems that protect approximately 8 million people from flooding face an escalating risk of failure under such conditions~\cite{qiu2023}. 

Quantifying the impacts of TCs on coastal infrastructure and vulnerable populations in a changing climate is essential for developing effective adaptation, mitigation, and resilience strategies. A critical aspect of this effort is assessing infrastructure robustness and resilience, which requires identifying and analyzing the most damaging TCs over long return periods. However, this task is challenging due to the limited historical record of TCs in many regions, like Bangladesh, and the non-stationary nature of climate change.

High-resolution numerical simulations are the standard approach for assessing future cyclone risks but are computationally prohibitive for large Monte Carlo ensembles. A statistical-physical framework that downscales TCs from coarse-resolution climate models is a promising alternative~\cite{emanuel2006statistical,emanuel2008hurricanes,ravela2010statistical}. This method parameterizes and reduces cyclone dynamics, enabling the rapid generation of sizeable synthetic cyclone ensembles. In Bangladesh, for example, this approach suggests that under a high greenhouse gas emission scenario, the likelihood of extreme cyclone winds exceeding 150 knots increases tenfold by the end of the century~\cite{emanuel_tropical_2021}. 

However, cyclone-induced hazards extend beyond extreme winds to include flooding, often the most damaging impact. Accurately modeling these hazards requires coupling downscaled wind fields with storm surge models and rainfall with inundation models for flooding (see Figure~\ref{fig:wkflw_before}). These simulations, which solve governing hydrodynamic equations, demand high computational resources, making it computationally infeasible to simulate the vast number of TCs required to identify the most damaging ones, see Figure~\ref{fig:wkflw_before}. For instance, assessing storm tide hazard in Bangladesh with ~5 climate scenarios, ~5 different SLR conditions, 12 global climate models, ~5,000 simulations per model to access the tails, and ~100 parameter perturbations across downscaling and hydrodynamic simulation will require $1.5\times10^8$ simulations to accurately assess storm tide risk for a single site in the world--an intractable computational burden.

\begin{figure}[htb!]
    \centering
    \includegraphics[width=0.5\textwidth]{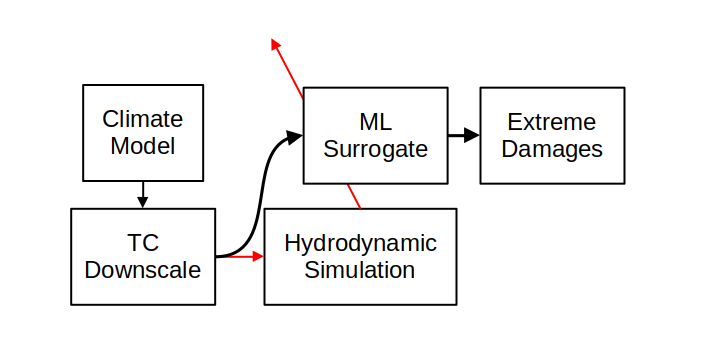}
    \caption{ML-based surrogate allows for faster risk assessment and search for damaging TCs. Few detailed simulations train the surrogate to locate the rare destructive storms, that is, those causing extreme storm tides, from a large number of downscaled TCs. }
    \label{fig:surrflow}
\end{figure}

To address these challenges, we propose a methodology that combines surrogate modeling with informative learning, see Figure~\ref{fig:surrflow}. Our approach efficiently identifies TCs likely to cause severe flood damage. It minimizes the need for computationally expensive hydrodynamic simulations while maintaining high precision. 

In this paper, we conduct a series of experiments to evaluate this methodology. In the first, a surrogate model trains on cyclone features as inputs to predict storm tides across 54 coastal locations in Bangladesh as outputs. We define damaging TCs as those producing water depths exceeding 3 meters above mean sea level at any location. Using thousands of synthetic TCs downscaled from ERA5~\cite{hersbach2020era5} reanalysis data for the 1981–2000 period. The surrogate model achieves a precision of 81\% on a 75\%-25\% training-test split. While this demonstrates skill in identifying damaging storms, the computational burden of simulating 75\% of an extensive training dataset remains a limitation.

In the second experiment, we test whether a surrogate model trained on ERA5~\cite{hersbach2020era5} data generalizes to a future climate scenario. Using TCs downscaled from the EC-EARTH-3~\cite{gmd-15-2973-2022} climate model under the SSP5-8.5~\cite{o2016scenario} scenario for 2081–2100, the surrogate model achieves a precision of 82\%, indicating its potential to generalize across different climates. In the third experiment, the training dataset is reduced to 20\% of the ERA5~\cite{hersbach2020era5} data while testing on the future climate scenario. Remarkably, the model maintains a precision of 80\%, demonstrating that accurate predictions are achievable with significantly fewer training simulations.

The fourth experiment employs an informative learning approach to optimize both data selection and model performance. Starting with a seed set of TCs representing just 1\% of the dataset, we train a surrogate model mapping storm parameters to storm tides. This iterative framework proposes new TCs likely to be damaging, and a small fraction of these are simulated in detail to retrain the surrogates for the subsequent iteration. By informatively coupling data selection with incremental model refinement, the active online learning system achieves 100\% precision after evaluating only 20\% of the dataset. This result highlights the method's ability to identify damaging TCs with minimal computational effort. Furthermore, the framework's flexibility enables prioritization of recall when desired, and its application extends to setting parameters for TC downscaling rather than relying on pre-generated samples, offering even greater efficiency.

These findings underscore the transformative potential of combining surrogate modeling with informative learning in cyclone hazard assessments. By achieving high precision with minimal computational cost, this methodology enables detailed risk assessments even in resource-constrained settings. The results open exciting possibilities for generalization to other climate scenarios, regions, and hazard types, offering a robust framework for advancing resilience and adaptation efforts in the face of intensifying cyclone risks.

The remainder of this paper is as follows. Section~\ref{sec:rw} surveys related work. Section~\ref{sec:methods} discusses pertinent details of downscaling and hydrodynamic simulation. Section~\ref{sec:batch} describes surrogate models based on batch learning, including climate scenario generalization. Section~\ref{sec:informative} presents details of the informative approach and associated experiments. A discussion follows in Section~\ref{sec:disc}, and the conclusions are presented with plans in Section~\ref{sec:concl}.

\section{Related Work}
\label{sec:rw}

The modeling and prediction of TCs, storm surges, and associated hazards have long relied on advanced numerical and statistical techniques. Statistical-physical frameworks, such as those introduced by Emanuel~\cite{emanuel2006statistical, emanuel2008hurricanes,ravela2010statistical}, enable efficient downscaling of TCs from coarse-resolution climate models to high-resolution representations, offering a rapid means to generate synthetic cyclone tracks and intensities. These foundational methods underpin many current cyclone risk assessments and have proven effective for generating large synthetic ensembles to study climate-driven changes in cyclone behavior. 

Coupling TCs with storm surge models has been a critical step in assessing coastal flooding risks~\cite{neumann2015risks}. While there are several models, we apply the ADCIRC model~\cite{luettich1992adcirc, westerink_adcirc_1994, pringle2021global} for simulating storm surges, demonstrated on events like Hurricane Katrina as well as simulations of flood risks in numerous applications~\cite{lin2012physically}. In Bangladesh, the focus area in this study, recent research efficiently evaluated storm surge risks in Bangladesh using Emanuel et al.'s synthetic TCs~\cite{khan2022storm}. However, this assessment is limited to the current climate, and to the best of our knowledge, Qiu et al.~\cite{qiu2023} provide the first quantitative assessments of climate change. 

There has been substantial interest in surrogate modeling for storm surges (and, by extension, storm tides). Jia et al.~\cite{jia2016} investigate Kriging for surrogate storm surge modeling utilizing an existing database of high-fidelity, synthetic storms. Kyprioti et al.\cite{KYPRIOTI2023104231} use two variations on Gaussian Processes (Kriging) with Principal Components. The first treats spatio-temporal predictions separably, leaving the storm features as the only surrogate model input. The second approach uses all sources to predict storm surges across space, time, and storm features. Zhang et al.~\cite{Zhang2018} 
describe an adaptive Gaussian Process regression framework (Kriging) using a sequential selection method that iteratively identifies storms that enhance prediction when added to the already available database. We refer the readers to Rasmussen and Williams~\cite{rasmussen2005gaussian} for an introduction to Gaussian process-based surrogate modeling. 

In contrast to these methods, we employ an XGBoost~\cite{Chen_2016} and ensemble-approximate conditional Gaussian Process (Ens-CGP)~\cite{saha2024,mirhoseini2024,trautner2020informative,Ravela2010,Ravela2007,Evensen2003} as the surrogate model. The latter utilizes an ensemble of downscaled TCs and their hydrodynamic simulations to determine the kernel without Kriging. 

Few studies exist that employ learning to enhance TC downscaling. For example,  Lu et al.\cite{Lu2024} developed parameterized cyclone precipitation models using random forest and extreme gradient boosting, achieving localized improvements in precipitation forecasting. In the context of hydrodynamic modeling, Zahura et al.~\cite{zahura2020training} show an example of training directly from high-fidelity models for flooding, and Pachev et al.~\cite{pachev2022framework} develop a surrogate model for peak storm surge prediction independently for each point of interest based on a multi-stage approach that first classifies points as inundated and then predicts the inundation level. We also directly predict the peak storm tide across a coastline, but not the time series and in a single step. The storm tide vector codes all the locations (1D vector), and we demonstrate generalization across climate scenarios. Further, we show an informative framework to learn the models with a few simulations and tune them for precision or recall. 

A key element of our approach is an incremental online adaptation of the surrogate model to achieve high precision or recall. The surrogate model iteratively improves data selection, while the selected data iteratively improves the surrogate model. This framework, to the best of our knowledge, has origins in Chernoff's sequential design~\cite{chernoff59} and Fedorov's optimal experiment design~\cite{fedorov1972theory} and its subsequent improvements, e.g., by Atwood~\cite{atwood73}. Active learning builds on these seminal ideas~\cite{cohn96a,cohn96b}, also see Settles~\cite{Settles2010}  for a review. However, unlike the use of an oracle, here we use a surrogate model and the expected information gain. Both XGBoost and Ens-CGP~\cite{saha2024} act as surrogates. Please see Schulz~\cite{schulz18} for a review of Gaussian Processes for surrogate modeling. The notion of informativeness appears in other literature, for example, adaptive obswervations~\cite{Morss2001}. Our informative approach follows prior work~\cite{trautner2020informative,Ravela2018}. Still, it has specific origins in Mackay's work~\cite{mackay92} and modifies the definition of informativeness in terms of destructiveness: the number of locations where the storm tide crosses the damage threshold.

\section{Methods and Data}
\label{sec:methods}
In this section, we briefly describe the generation of simulation-based training data. It includes a discussion of TC downscaling, hydrodynamic modeling, and the current and future climate scenarios for assessing surrogate performance. 

\subsection{Tropical Cyclone Downscaling}

We use a statistical-deterministic downscaling technique to create sets of synthetic TCs that affect Bangladesh~\cite{emanuel2006statistical,emanuel2008hurricanes}. The method uses thermodynamic and kinematic statistics from gridded global reanalyses or climate models to produce many synthetic TCs. Initially, we synthetically generate wind time series at 250 and 850 hPa levels, each as a Fourier series of random phases in time with a geostrophic turbulence power-law distribution of the kinetic energy spectrum, together constrained to have accurate monthly means, variances, and covariance from the reference climate model. The weighted average of synthesized winds according to the beta-and-advection model~\cite{holland1982tropical} synthesizes storm tracks. The time-evolving environment is seeded randomly in space and time with warm-core vortices drawn from a Gaussian distribution of peak wind speeds centered at 12 m/s (25 knots). These seed vortices then propagate forward. Following the track, the intensity of the vortices is then calculated deterministically using the Coupled Hurricane Intensity Prediction System (CHIPS) model~\cite{emanuel2004environmental}, which phrases the dynamics in angular momentum coordinates that allow for very high spatial resolution in the storm core. The thermodynamic input to the intensity model includes monthly mean potential intensity, along with 600 hPa temperature and specific humidity, derived from global climate models. Additional large-scale environmental factors such as potential intensity, wind shear, humidity, and ocean thermal stratification come from gridded global reanalyses or climate models following the track, controlling the dynamics of the synthetic cyclone. The storms used here are identical to the ones used in the cyclone study~\cite{emanuel_tropical_2021}.

Over 99\% of the seeded tracks dissipate quickly and are discarded. The remaining successfully grow to make up the downscaled TC climatology of a reanalysis or climate model. Only seeds that reach a maximum wind speed of at least 21 m/s (40 kt) during their lifetime form synthetic TCs, and each simulated synthetic TC is an hourly time series of storm parameters, including time, central position, maximum wind speed, pressure deficit, and radius to maximum wind. We identify synthetic TCs affecting Bangladesh based on their passage over coastal line segments~\cite{emanuel_tropical_2021}. Here, we use the reanalysis (ERA5~\cite{hersbach2020era5}) and EC-EARTH-3~\cite{gmd-15-2973-2022}  climate models. Please see Emanuel et al.~\cite{emanuel2008hurricanes} for comparisons of downscaled TC behavior with observations across basins. 

\subsection{Hydrodynamic Modeling of Storm Tides}
Here, we provide a brief description of the model setup, following~\cite {qiu2023}, which we refer the reader to for details. We used ADCIRC (ADvanced CIRCulation model, two-dimensional barotropic tides, Version 55.01)~\cite{luettich1992adcirc, westerink_adcirc_1994, pringle2021global} for storm surge simulations and used a tool called OceanMesh2D, to generate detailed, high-fidelity unstructured meshes~\cite{roberts2019oceanmesh2d, roberts2019automatic} for Bay of Bengal (spanning latitudes from 9°N to 23°N and longitudes from 80°E to 100°E). The final unstructured mesh consisted of 62,009 vertices and 115,199 triangular elements, with a resolution ranging from 20 km over the deep ocean to 1 km near the coastlines. The model accounts for all eight major astronomical tidal components ($K_1$, $K_2$, $M_2$, $N_2$, $O_1$, $P_1$, $Q_1$, and $S_2$). 

The surface wind and atmospheric pressure field associated with a TC is reconstructed at each node using the symmetric Holland parametric vortex model (H80) during the simulation~\cite{holland1980analytic}. A spatially varying \textit{Manning's N} parameterizes bottom friction. In simulation, to balance computational cost and numerical stability, a time step of 60 seconds is used. A simulation for one track takes approximately 8.8 seconds using a parallel setup with 40 CPU cores. A one-day model spin-up applies to all simulations. 

Astronomical tides synchronize with synthetic TC timing, and we assess the hydrodynamic model's performance by comparing its output with TPXO9-Atlas and tide gauge station data for global astronomical tide validation and total water level validation. In comparison with previous studies~\cite{sindhu2013characteristics, rose2022tidal}, the model accurately describes the constituents' general response, including the amphidromes' positions. Subsequently, the historical TC Sidr (``IO062007") was used to validate the total water levels. The results, including the overall root-mean-square error, bias, and Willmott skill at four tidal stations, show satisfactory agreement between observed and simulated storm tides. Thus, we ensure that the hydrodynamic model is suitable for capturing storm surge dynamics and water level variations in coastal Bangladesh. However, please note that for the simulations used in this paper, we do not incorporate sea level rise, and references to water level are with respect to mean sea level.

\subsection{Training and Testing Data Set}

For developing the surrogate model, we use datasets of synthetic TCs downscaled from the ERA5~\cite{hersbach2020era5} reanalysis for the present climate and the EC-EARTH-3~\cite{gmd-15-2973-2022} SSP5-8.5~\cite{o2016scenario} to represent a future climate scenario. Our ERA5~\cite{hersbach2020era5} dataset consists of 4100 synthetic storm tracks, and our SSP5-8.5~\cite{o2016scenario} dataset consists of 2000.

The data for each synthetic TC track in consideration comes in a time series that consists of $> 100$ 1-hour time steps. As input data into the XGBoost~\cite{Chen_2016} model, we extract 33 of these time steps for each TC: the 24 hours before landfall, the landfall hour, and the 8 hours after landfall. If a TC never reaches land or has two or more landfalls, we exclude it from our experiments. For each TC, at each selected time step, we consider five out of the many parameters for input into the model: latitude, longitude, maximum wind speed (VMAX), minimum sea level pressure (MSLP), and radius of maximum wind (RMW).  

We read the hydrodynamic model output along the coastline at 54 ``virtual" stations (see~\cite{qiu2023}). At each of these stations, the peak total water depth provides entries to a 54-element vector. Thus, the ML model should output an array of 54 values for each TC, each representing the maximum storm tide in meters throughout the storm track at one virtual station along the coastline of Bangladesh. 

\section{Learning Batch Surrogate Model}
\label{sec:batch}

We selected XGBoost~\cite{Chen_2016} as the surrogate model after several attempts at other models that did not perform as well. In the paper's experiments, the XGBoost parameters are as follows: gbtree booster, eta = 0.3, gamma = 0, max\_depth = 6, min\_child\_weight = 1, max\_delta\_step = 0, subsample = 1, sampling\_method = uniform, lambda = 1, alpha = 0, tree\_method = hist, scale\_pos\_weight = 1, refresh\_leaf = 1, process\_type = default, grow\_policy = depthwise, and max\_bin = 256. 
In this section, we train an XGBoost surrogate model directly, testing its training data needs and its ability to generalize to new climate scenarios and models.

\subsection{Experiment I: ML can predict storm surges in present climate}

Using a random split of the ERA5~\cite{hersbach2020era5} dataset with 75\% of storm tracks used for training and the remaining 25\% used for testing, we assess the ability of ML to predict storm surges in the present climate. We repeat 50 trials of this process, using a different random split of 75\%/25\%, training and testing a new model each iteration. 

\begin{figure}[h]
    \centering
    \includegraphics[width=0.4\textwidth]{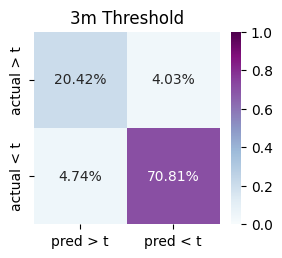}
    \caption{Average confusion matrix from repeated trials of training model on a random split of 75\% of ERA5~\cite{hersbach2020era5} data and testing on the remaining 25\%.}
    \label{fig:exp1}
\end{figure}

We will call a TC that produces a surge of $>3$ meters at any of the Bangladesh coastline stations as a ``positive", damaging storm, and a storm with no surge $>3$ meters at any station as a ``negative", not destructive storm. We categorize the TCs into $>3$m and $\leq 3$m categories based on the predictions of the ML model and draft a confusion matrix of this classification (true positives in top left, true negatives in bottom right, false positives in bottom left, and false negatives in top right). Figure~\ref{fig:exp1} shows the average confusion matrix values over the 50 trials. 

The averaged model predictions have a recall of 84\%, a precision of 81\%, an accuracy of 91\%, and a misclassification of 9\%. Please note that recall is the proportion of damaging storm tides retrieved, precision is the proportion of retrieved damaging storm tides, and accuracy is the proportion of correctly classified damaging and non-damaging storm tides.

Thus, XGBoost appears to predict whether a storm in the present climate is destructive or not with high accuracy, precision, and recall. Using a model trained on synthetic TCs from our present climate, we can predict whether a new, unseen storm is destructive. However, there are at least two concerns. First, we do not know if the algorithm can generalize to a new climate model and scenario, and second, we don't understand how data-hungry the approach is. We will test these two aspects in the following two experiments.

\subsection{Experiment II: ML can predict storm surges in a future climate scenario}

\begin{figure}[h]
    \centering
    \captionsetup{justification=centering}
    \includegraphics[width=0.4\textwidth]{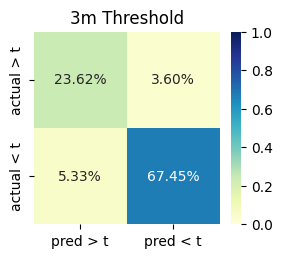}
    \caption{Confusion matrix of results from training model on all ERA5~\cite{hersbach2020era5} data and testing on all SSP5-8.5 data~\cite{gmd-15-2973-2022,o2016scenario}.}
    \label{fig:exp2}
\end{figure}

For this experiment, we train an XGBoost~\cite{Chen_2016} model (with the same parameters) on the full ERA5~\cite{hersbach2020era5} dataset and test its ability to predict storm surges for all of the EC-EARTH-3~\cite{gmd-15-2973-2022} SSP5-8.5~\cite{o2016scenario} storm tracks. Figure~\ref{fig:exp2} plots the test predictions for the SSP5-8.5~\cite{o2016scenario} dataset as a confusion matrix. 

The model predictions have a recall of 87\%, a precision of 82\%, an accuracy of 91\%, and a misclassification of 9\%. Thus, the ML model trained on TCs produced in the present climate can be used to accurately predict the destructiveness of TCs produced in a future climate scenario. There are, however, a few caveats that Section~\ref{sec:disc} discusses.

\subsection{Experiment III: ML needs very little data to be skillful}
\begin{figure}[h]
    \includegraphics[width=0.4\textwidth]{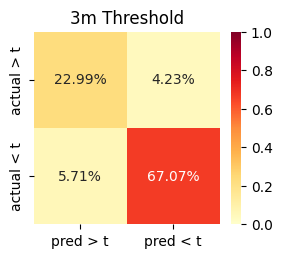}
    \caption{Averaged confusion matrix from repeated trials of training model on a random split of 20\% of ERA5~\cite{hersbach2020era5} data and testing on all SSP5-8.5 data~\cite{gmd-15-2973-2022,o2016scenario}.}
    \label{fig:exp3}
\end{figure}

Training the XGBoost~\cite{Chen_2016} model on a random split of 20\% of ERA5~\cite{hersbach2020era5} data and testing on all of SSP5-8.5 data~\cite{gmd-15-2973-2022,o2016scenario}, we obtained similar results to Experiment II, where we trained on all of ERA5~\cite{hersbach2020era5} data before testing. We repeat 50 trials of the train-test process, using a different random split of 20\% for training, training, and testing a new model each iteration. Figure~\ref{fig:exp3} plots the average of test predictions as a confusion matrix. 

The averaged model predictions have a recall of 84\%, a precision of 80\%, an accuracy of 90\%, and a misclassification of 10\%. We see that when tested on SSP5-8.5~\cite{o2016scenario} data, an ML model that only sees 20\% of ERA5~\cite{hersbach2020era5} data in training can perform nearly as well as a model that considers all of the data. Thus, training requires fewer hydrodynamic simulations. Still, we would like to understand better if the precision (or recall) can be improved using only a few training data. The following section addresses this issue.

\begin{figure}[htb!]
    \includegraphics[width=0.5\textwidth]{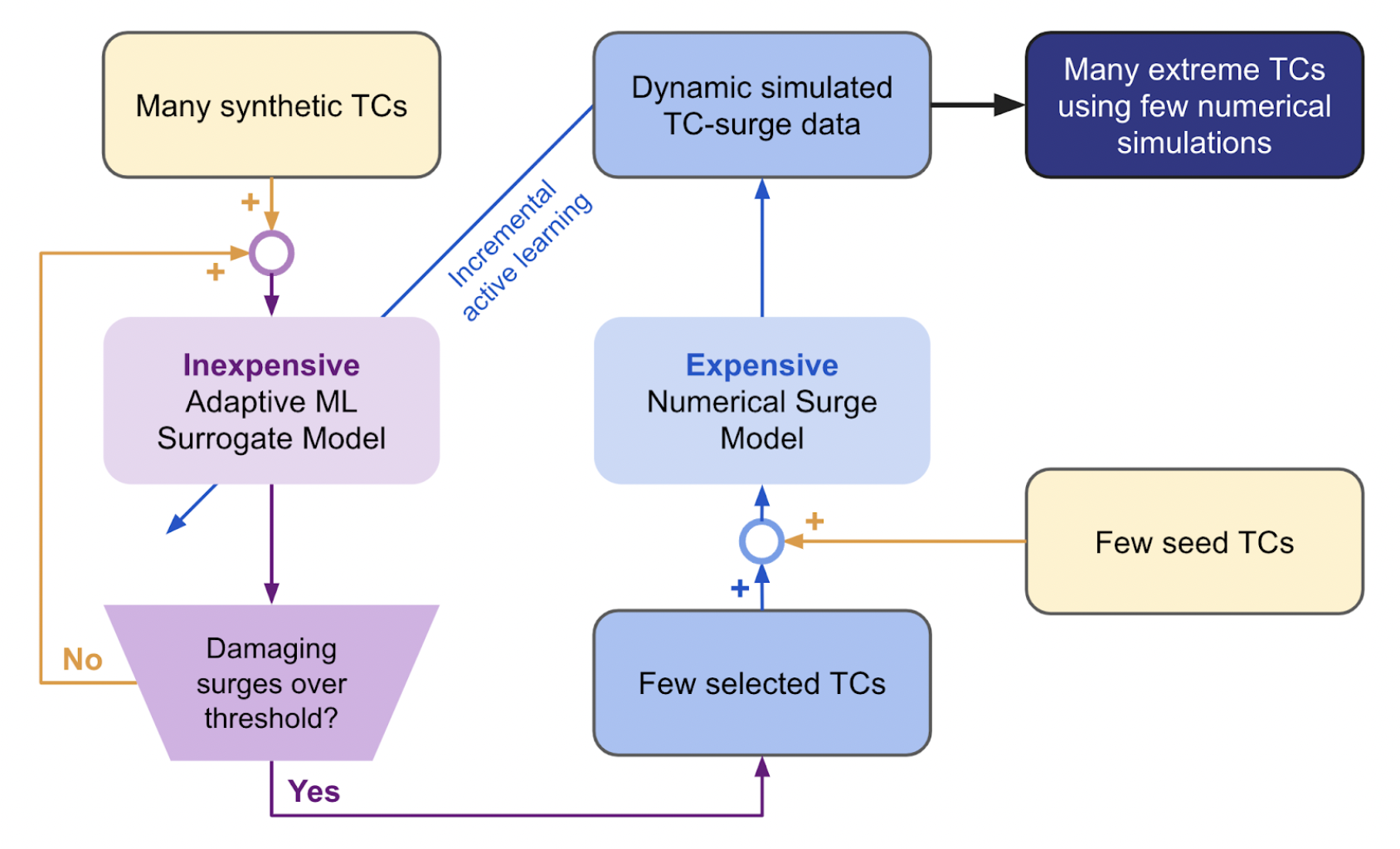}
    \caption{The flowchart for active and informative sampling to search for extreme storm tides with few hydrodynamic simulations. It requires iterative surrogate refinement paired with data selection, which we test with and without informative selection strategies.  }
    \label{fig:exp4_diagram}
\end{figure}

\section{Informative Learning Accelerated Search}
\label{sec:informative}

Informative learning~\cite{trautner2020informative} refers to an approach that maximizes expected information gain by selecting models, data, experts, or constraints~\cite{Ravela2018}. Such closed-loop systems continually use feedback between these processes to deliver extraordinary predictive or discovery solutions. Our primary objective is to predict storm tides using a few detailed simulations. In the informative approach, shown in Figure~\ref{fig:exp4_diagram}, we actively filter a large number of storm simulations using a surrogate model, selecting a small subset of training data likely to produce damaging water depths. This small, filtered subset of TCs is then numerically simulated, refining the surrogate model online. Incrementally, the concentration of destructive TCs and the surrogate model's ability to locate damaging TCs improve. Because each iteration involves only a few numerical simulations, either precision or recall is attained quickly.

In our application, we aim for informative sampling to provide total precision, increasing the proportion of damaging TCs approaching one. When picking from a large but finite set of TCs, however, precision often comes at the cost of recalling only a few simulated TCs at a time. In situations where high precision is paramount, this approach is ideal. In other circumstances, where the goal is to locate all synthetic TCs generating damaging storm tides, the high recall may be prioritized, albeit at the expense of precision. Our approach is tunable to meet either requirement.

The surrogate model forms the key element of the informative approach. Below, we describe two different classes of models in separate experiment subsections.

\begin{figure*}[htb!]
    \centering
    \captionsetup{justification=centering}
    \includegraphics[width=0.95\textwidth]{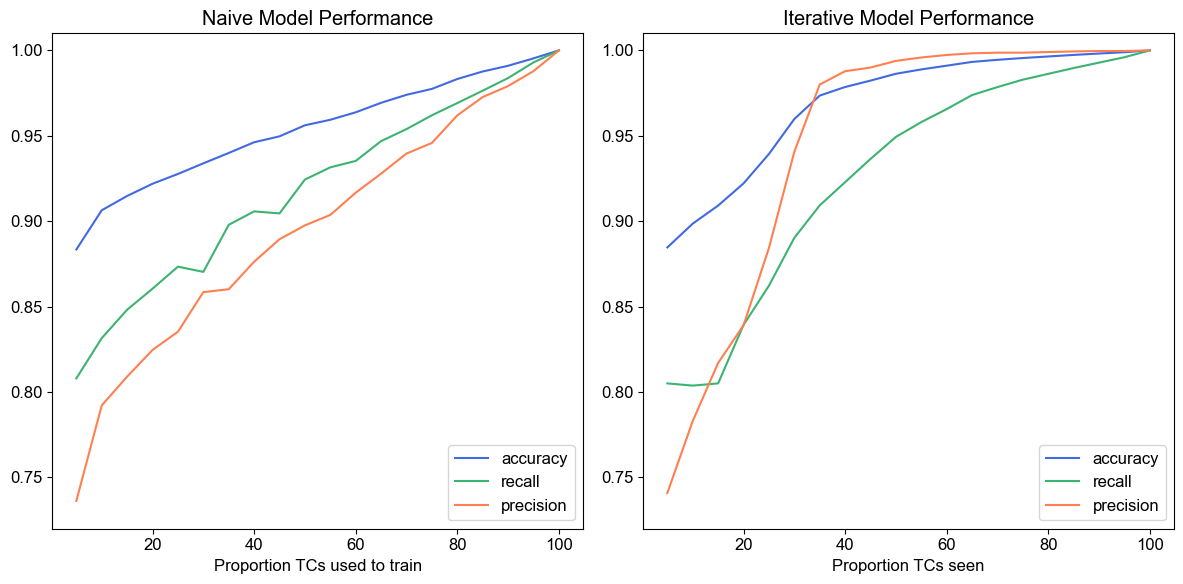}
    \caption{Comparison of naive and iterative approaches for storm tide prediction using the XGBoost surrogate model. The iterative process incrementally improves model performance by concentrating on destructive TCs that cross the threshold without further prioritizing destructiveness.}
    \label{fig:exp4_graphs}
\end{figure*}

\subsection{Experiment IVa (XGBoost): Iterative ML Model System Reduces Data Requirements}

The first surrogate model explored is XGBoost~\cite{Chen_2016}, with previously set parameters. Using a few initial seeds, $1\%$ of TCs, we feed these into the computationally expensive hydrodynamic numerical model to create a small training dataset. This data pool trains the initial XGBoost ML model. An iterative closed-loop process then adapts the surrogate model.

From the more extensive set of hydrodynamically unsimulated TCs, the surrogate model predicts storm tides at each of the 54 stations. Filtering based on the number of destructive surges exceeding the 3m threshold, we select the top $5\%$ of TCs and simulate them numerically. These simulations expand the training dataset and increase the concentration of destructive TCs. The updated training data retrains the surrogate ML model, and the loop repeats. Figure~\ref{fig:exp4_diagram} illustrates this process.

Figure~\ref{fig:exp4_graphs} shows that the naive strategy—training with an increasing fraction of TCs and their numerically simulated surges while testing on the decreasing fraction of remaining storms—lacks efficacy. Recall, precision, and accuracy all grow nearly linearly. In contrast, the closed-loop informative process with incremental online adaptation achieves very high precision without requiring a large number of storms. These graphs were the average of results over 10 trials for each training strategy. While the results demonstrate promise, we seek to improve both recall and precision further. The following section describes a second informative approach using an incrementally and online-trained surrogate model.

\subsection{Experiment IVb (Ens-CGP): Iterative ML model for perfect precision or total recall}

The second surrogate model is an ensemble-approximated conditional Gaussian process (Ens-CGP)~\cite{saha2024,yamaguchi2023,trautner2020informative,Ravela2007}. This model predicts storm tide vectors based on TC parameters. Let the input vector $\underline{x} \in \mathbb{R}^N$ ($N = 165$) represent TC parameters, and the output vector $\underline{y} \in \mathbb{R}^O$ ($O = 54$) represent the maximum recorded storm tide at coastal locations. Akin to the EnKF~\cite{Evensen2003}, the Ens-CGP~\cite{saha2024} model is defined as:

\begin{equation}
    M = C_{yx}\left[C_{xx}+\sigma^2 I\right]^{-1},
\end{equation}

where $C_{xx}$ is the covariance of storm parameters, $C_{yx}$ is the cross-covariance between storm tide vectors and storm parameters, and $C_{yy}$ is the covariance of storm tides across the ``virtual" stations along the coastline. The scalar $\sigma^2$ is a ridge-inducing regularization parameter chosen empirically. The model predicts as  follows:

\begin{equation}
    \hat{y} = \underline{\bar{y}} + M (\underline{x} - \underline{\bar{x}}),
\end{equation}

where $\underline{\bar{x}}$ and $\underline{\bar{y}}$ are the mean storm parameter vector and storm tide vector, respectively, derived from the training data (initially, a randomly chosen $1\%$ seed sample of TCs). The input parameter vector is $\underline{x}$, and the estimated storm tide vector is $\underline{\hat{y}}$. 

\subsubsection{Training Procedure}

Instead of explicitly calculating the covariances, a reduced-rank square-root ensemble representation provides computational efficiency. Given a training dataset $\triangle_j = \{(\underline{x}_1, \underline{y}_1), \ldots, (\underline{x}_{E_j}, \underline{y}_{E_j})\}$ at the $j^{\text{th}}$ retraining iteration, the following steps are performed:

\begin{enumerate}
    \item \textbf{Calculate Training Ensemble Means:}
    \begin{eqnarray}
        \underline{\bar{x}}_j &=& \frac{1}{E_j}\sum_{i=1}^{E_j} \underline{x}_i, \\
        \underline{\bar{y}}_j &=& \frac{1}{E_j}\sum_{i=1}^{E_j} \underline{y}_i.
    \end{eqnarray}
    
    \item \textbf{Compute the Deviation Ensemble Matrices:}
    \begin{eqnarray}
        \tilde{X}_j &=& \left[\underline{x}_1 - \underline{\bar{x}}_j, \ldots, \underline{x}_{E_j} - \underline{\bar{x}}_j \right], \\
        \tilde{Y}_j &=& \left[\underline{y}_1 - \underline{\bar{y}}_j, \ldots, \underline{y}_{E_j} - \underline{\bar{y}}_j \right].
    \end{eqnarray}
    
    \item \textbf{Perform Singular Value Decomposition (SVD):}
    \begin{equation}
        \tilde{X}_j = U \Sigma V^T.
    \end{equation}
    
    \item \textbf{Estimate the Reduced Ens-CGP Model:}
    \begin{equation}
        M_j = \tilde{Y}_j V_\beta \Sigma_\beta \left[\Sigma_\beta^2+\sigma^2 I\right]^{-1} U_\beta^T,
    \end{equation}
    where $U_\beta = U(:, 1:k)$, $V_\beta = V(:, 1:k)$, and $\Sigma_\beta = \Sigma(1:k, 1:k)$. The selected value of $k$ covers $99\%$ of the cumulative energy of the singular values, ensuring $k < \text{Rank}(\Sigma)$.
\end{enumerate}

\subsubsection{Iterative Refinement Process}

At iteration $j$, the model $M_j$ predicts storm tide vectors for storm parameter vectors not in the training dataset. From these predictions, TCs exceeding the damage threshold (e.g., $> 3$ meters) are ranked based on their destructiveness. the training iteration is updated ($j \leftarrow j+1$) after simulating the top $1\%$ of the most damaging surrogate-classified TCs numerically.

The initial surrogate models may misclassify TCs, predicting non-damaging storms as damaging and vice versa. However, by aggressively selecting the most destructive TCs, the training data becomes increasingly concentrated, improving model performance with each iteration. The damage threshold is adjustable to bias the sampler toward precision or recall. Relaxing the threshold enables the model to prioritize specific performance metrics but induces more false positives or false negatives.

\begin{figure}[htb!]
    \centering
    \includegraphics[width=0.9\linewidth]{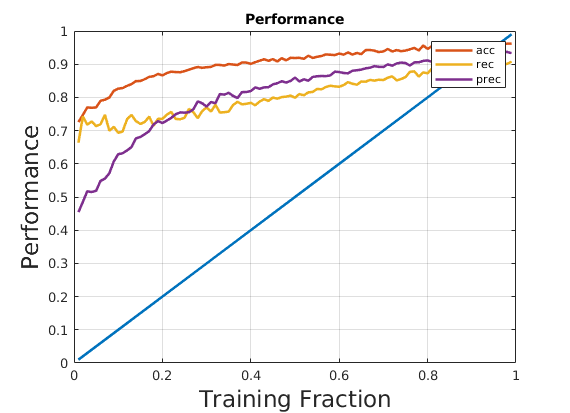}
    \includegraphics[width=0.9\linewidth]{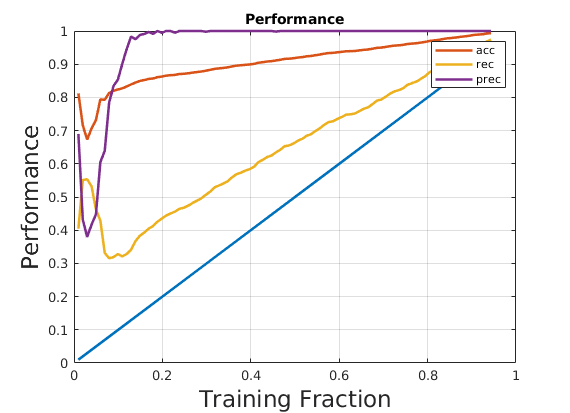}
    \includegraphics[width=0.9\linewidth]{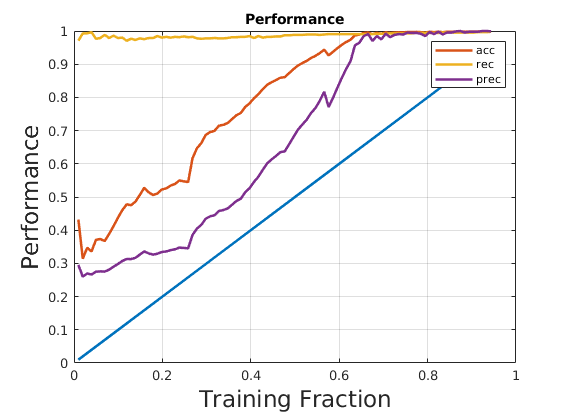}
    \caption{Comparison of naive batch training (top), precision-tuned informative sampling (middle), and recall-tuned informative sampling (bottom) using the Ens-CGP surrogate model. The precision-tuned sampler of interest in extreme risk achieves $100\%$ precision early in the process while maintaining progress toward total recall. This approach quantifies and prioritizes the destructiveness of surrogate-model predicted storm tides.}
    \label{fig:coact}
\end{figure}

Figure~\ref{fig:coact} shows the three cases where Ens-CGP informatively selects the top 1\% TCs by destructiveness, untuned, and tune for precision or recall. Near-perfect performance emerges in the latter case with little data. 

\section{Discussion}
\label{sec:disc}

The batch approach, which trains on only 20\% of the data, demonstrates high recall, precision, and accuracy. This result suggests that surrogate models are not only effective for data selection but can also play a critical role in search procedures for identifying extreme events. The generalization from the current climate to the SSP5-8.5~\cite{o2016scenario} scenario is particularly promising, though a few caveats exist. Firstly, the testing TC tracks inherently reflect the SSP5-8.5~\cite{o2016scenario} future climate scenario, embedding some of its characteristics—such as higher wind speed return periods—into the parameter space. Secondly, the hydrodynamic simulations assume a constant sea level for all storm tracks, including those representing future climates. Consequently, we evaluate the machine learning model's predictions under the simplifying assumption of zero sea-level rise. While this limitation affects the robustness of the generalization claim, it is worth noting that one can incorporate sea-level rise as a post-processing step~\cite{qiu2023}. Despite these caveats, the results remain encouraging for climate-related applications.

The active learning approaches incrementally refine the surrogate model online by using it for data selection and subsequently updating the model with the selected data~\cite{Ravela2018,trautner2020informative}. This iterative process continually improves the surrogate model's performance, focusing on TCs that achieve the targeted damage threshold. However, the two active learning methods differ significantly in their execution. The first method employs XGBoost as the surrogate model and randomly samples a fraction of the predicted storm tide vectors without specifically evaluating information gain in its sequential experiments~\cite{chernoff59} or experimental design~\cite{fedorov1972theory}. The second method uses an ensemble-approximated conditional Gaussian process (Ens-CGP), quantifying information gain by identifying the most informative (damaging) TCs for iterative experiment design. Without access to an oracle~\cite{Settles2010}, this approach exemplifies informative learning~\cite{trautner2020informative,mackay92} using the surrogate. Furthermore, while the XGBoost method does not modify the damage threshold, the Ens-CGP approach varies this threshold to emphasize either recall or precision. The significant reduction in simulations achieved by the informative approach underscores its potential for practical applications.

Conceptually, the surrogate modeling process implements the forward Kolmogorov process, where an ensemble of storm parameter distributions produces an ensemble of storm tide vector distributions. For this study, which focused on data selection and model refinement, this forward modeling suffices. However, a more comprehensive implementation would require modeling both the forward and backward Kolmogorov processes. Such an approach would enable the use of errors between the predicted storm tide ensemble and the targeted damage distribution to refine the storm parameter distribution. This refinement could include assessing parameter uncertainties and leveraging the information they contain. By doing so, the method could guide subsequent TC simulations in a continuous generative framework rather than relying on selection from a fixed, extensive catalog. This expanded capability would offer greater flexibility and accuracy in the simulation and prediction of extreme storm events.

\section{Conclusions and Future Work}
\label{sec:concl}
This paper presents two surrogate modeling approaches to efficiently identify TCs that generate destructive storm tides with minimal reliance on detailed hydrodynamic simulations. Training data from TCs downscaled 
(ERA5~\cite{hersbach2020era5} and EC-EARTH-3~\cite{o2016scenario,gmd-15-2973-2022}) datasets using a statistical-physical method~\cite{emanuel2006statistical,emanuel2008hurricanes,ravela2010statistical,emanuel_tropical_2021}. ADCIRC provides hydrodynamic simulations incorporating tides but not sea-level rise to provide storm tide estimates for training and validation.

The first approach uses a batch training strategy, leveraging a fraction of the storm catalog generated from simulations. We demonstrate that the batch surrogate model generalizes well to new climate scenarios and achieves high recall, precision, and accuracy with a relatively small fraction of the dataset. The second approach employs an active, iterative data selection process to adapt surrogate models online. The approach explores two methods:
\begin{itemize}
\item XGBoost-Based Surrogate Model: Here, the data selection mechanism randomly samples TCs predicted to exceed the specified storm tide threshold, iteratively refining the surrogate model.
\item Informative Sampling Method: This method quantifies information gain as destructiveness, selecting a subset of the most destructive TCs for refinement. The approach achieves either perfect recall or precision after training with only a tiny fraction of the storm catalog and a variable thresholding mechanism.
\end{itemize}

The experiments suggest a scalable, generalizable, and efficient approach to sampling rare extremes. Although the results are for Bangladesh and a specific climate scenario, we posit that they are applicable to other regions and scenarios with minor modifications. 

This work has several future directions. First, we aim to thoroughly compare surrogate modeling approaches, including deep learning architectures, to evaluate their relative performance. We also plan to explore additional informative sequential experimental design regimes. The thresholding mechanism used to balance recall and precision is ad hoc, and reinforcement learning techniques hold promise for optimizing this policy. Another area of interest involves further improving the methodology by leveraging the forward-backward Kolmogorov process to enhance model adaptation.

The informative approach described here has potential applications beyond storm tide prediction. For example, in recent work~\cite{mirhoseini2024}, our group developed surrogate models for inundation prediction. The present approach could apply to select rainfall events that result in extreme inundation, and we plan to investigate this in future research.

\section*{Acknowledgment}
        This research is part of the MIT Climate Grand Challenge Weather and Climate Extremes and Jameel Observatory CREWSNet projects. Funding from Schmidt Sciences, LLC, Liberty Mutual (029024-00020), and ONR (N00014-19-1-2273) also supported this project.

\bibliography{refs,2110.11390,mainbib}
\bibliographystyle{plain}
\end{document}